\documentclass[multphys,vecphys]{svmult}

\usepackage{makeidx}     
\usepackage{graphicx}    
\usepackage{multicol}    

\makeindex             

\newcommand{\Msun}{~M_\odot}

\begin{document}

\newcommand{\lsun}{L_\odot}
\newcommand{\gcm}{\rm ~g~cm^{-3}}
\newcommand{\cmc}{\rm ~cm^{-3}}
\newcommand{\kms}{\rm ~km~s^{-1}}
\newcommand{\ergs}{\rm ~erg~s^{-1}}
\newcommand{\ergcc}{\rm ~erg~cm^{-3}}
\newcommand{\ml}{~\Msun ~\rm yr^{-1}}
\newcommand{\mll}{\Msun ~\rm yr^{-1}}
\newcommand{\Mdot}{\dot M}
\newcommand{\la}{\raise0.3ex\hbox{$<$}\kern-0.75em{\lower0.65ex\hbox{$\sim$}}}
\newcommand{\ga}{\raise0.3ex\hbox{$>$}\kern-0.75em{\lower0.65ex\hbox{$\sim$}}}

\title*{The Surroundings of Gamma-Ray Bursts: \\
Constraints on Progenitors}
\author{Roger A. Chevalier}
\institute{Department of Astronomy, University of Virginia, P.O. Box
3818, \\
Charlottesville, VA 22903, USA
\texttt{rac5x@virginia.edu}}
%
%
\maketitle

\begin{abstract}
The association of a supernova with a gamma-ray burst (GRB 030329)
implies a massive star progenitor, which is expected to have an
environment formed by pre-burst stellar winds.
Although some sources are consistent with the expected wind environment,
many are not, being better fit by a uniform density environment.
One possibility is that this is a shocked wind, close to the burst
because of a high interstellar pressure and a low mass loss density.
Alternatively, there is more than one kind of burst progenitor, some
of which interact directly with the interstellar medium.
Another proposed environment is a pulsar wind bubble that has
expanded inside a supernova, which requires that the supernova
precede the burst.

\end{abstract}

\section{Introduction}
\label{sec:1}

Some of the best evidence for nature of gamma-ray burst (GRB)
progenitors has come from the identification of the Type Ic
supernovae SN 1998bw with GRB 980425 \cite{Ge98}
and of the recent SN 2003dh with GRB 030329 \cite{Se03}.
The finding of these events supports models of long-duration
GRBs originating from stripped massive stars \cite{MWH01}.
The surroundings of massive stars are expected to be shaped by
the winds emanating from the progenitor stars.
Clear evidence for the signature of a wind has been difficult to
establish, and the possibility remains that
there is more than one type of progenitor for the long-duration
bursts.
Direct interaction with the interstellar medium might be
expected if the progenitor involves a compact binary system.

Another possibility is that the GRB occurs in a massive star
weeks to years after it has become a supernova \cite{VS98}.
The progenitor of the
GRB may be a rapidly spinning neutron star that spins down and
eventually collapses, leading to a burst.
This can create a pulsar wind nebula immediately surrounding
the burst progenitor, which has possible advantages for producing
a GRB \cite{KG02}.

\section{Afterglows}

\subsection{Afterglows and the circumburst medium}

The afterglows of GRBs provide a probe of the immediate surroundings
of GRBs.
The evidence that we have from bursts related to supernovae is that
the progenitors are Type Ic supernovae.
In addition to SN 1998bw and SN 2003dh, the burst GRB 021211 has possibly
been identified with a Type Ic supernova like SN 1994I \cite{DVe03}.
These supernovae are thought to have massive star progenitors that
have been stripped of their hydrogen envelopes, i.e. Wolf-Rayet
stars.
This type of massive star progenitor is also suggested by the argument
that the collimated flow from a burst be able to pass through
the star in a time that is not significantly longer that the duration
of the GRB.
Even for the long duration bursts, this implies a relatively compact
stellar progenitor like the Wolf-Rayet stars
\cite{MWH01,M03}.

Wolf-Rayet stars are found to have winds with typical mass loss rate
$\dot M= 10^{-5}\ml$ and wind velocity $v_w=1000\kms$ ending in a
termination shock where the wind runs into the surrounding medium
\cite{CL99}.
A steady wind produces a density distribution $\rho=A r^{-2}$;
the value of the density can be scaled to the corresponding
value for the standard wind
 parameters, $A_*=A/(5\times 10^{11} {\rm~ gm~cm^{-1}})$.
For a surrounding medium of pressure $p$, the wind termination shock occurs
at a radius
\begin{equation}
R_t=5.7\times 10^{19}\left(v_w\over 1000\kms\right)
\left(p/k\over 10^4~{\rm cm^{-3}~K}\right)^{-1/2} A_*^{1/2} 
{\rm~cm}
\end{equation}
where $k$ is Boltzmann's constant.
The pressure is normalized to a typical value in the interstellar medium of
our Galaxy.
A higher pressure can occur as a result of the wind bubble evolution, or
as the result of an especially high pressure interstellar medium.

For a spherical explosion with energy $E$, the blast wave 
in the free wind reaches the
radius $R_t$ after a time
\begin{equation}
t_t=1.35\times 10^4\left(v_w\over 1000\kms\right)^2
\left(p/k\over 10^4~{\rm cm^{-3}~K}\right)^{-1}
\left(E\over 10^{53}{\rm~ergs}\right)^{-1} A_*^{2} {\rm~days}.
\label{time}
\end{equation}
This shows that over typical times of observation, the blast wave may
be expanding into the free wind for standard parameters.
In view of this, interaction with a wind has become one of the 
models that is investigated in modeling the afterglows of GRBs \cite{CL00}.
This model is compared to results for a constant density medium, which
was initially taken in afterglow modeling as the simplest assumption.
Taking the surrounding density of the form $\rho\propto r^{-s}$,
these cases can be designated $s=0$ (uniform) and $s=2$ (wind).
In afterglow models, 
both wind and constant density models provide adequate fits
to the data
in some cases, although the constant density model is usually favored
 \cite{PK01,PK02}.

At first sight, this ambiguity is surprising.
In the context of models with constant efficiencies evolving before
a jet break sets in, the two cases have distinctive behavior:
the synchrotron self-absorption frequency, $\nu_a$, evolves to
higher frequency with $s=2$, but remains constant for $s=0$;
the peak flux drops with time for $s=2$, but remains constant for $s=0$;
the synchrotron cooling frequency, $\nu_c$,  
evolves to higher frequency for $s=2$,
but evolves to lower frequency for $s=0$.
There are various reasons why these differences have not provided clear
tests of the models.
The evolution of $\nu_c$ requires good light curve 
information at optical/IR and
X-ray wavelengths, which is usually lacking.
The evolution of $\nu_a$ requires early radio data; 
these data are usually sparse
and are affected by interstellar scintillation.
Another problem is that jet breaks are observed in light curves 
and the evolution
in the post-break regime can mimic some of the features of evolution in
an $s=2$ medium: the peak flux drops with time and $\nu_a$ evolves to higher
frequency \cite{SPH99}.

Another possibility for distinguishing between the models is to go to very
early times, within about a minute of the GRB burst.
In addition to the older case of GRB 990123, this has recently been
achieved for GRB 021004 \cite{Fox03} and GRB 021211 \cite{Le03,We02,Pe02}.
The advantage of these early times is that jet effects do not play
a role in the evolution.
Li \& Chevalier \cite{LC03} suggested that the early flat optical light curve
of GRB 021004 could be interpreted in terms of wind interaction, in which
the critical frequency  $\nu_m$ had not yet moved down through
optical wavelengths.
Although this model has some promise, the case remains ambiguous in that
the relatively flat evolution might also be produced by a combination of
emission from the reverse shock wave and the later forward shock emission
in the case of interaction with a $s=0$ medium \cite{KZ03}.
To distinguish between these possibilities, both good light curve data
and color information are needed.

An additional problem with GRB 021004 is that the optical light curve
showed variability superposed on the overall trend \cite{He03}.
This makes it difficult to clearly specify characteristic times in the
evolution of the afterglow.
The variability, which is not seen in all afterglows, may be due to
density inhomogeneities in the circumburst medium and may thus provide
additional diagnostics for that medium \cite{HP03}.
The winds of Wolf-Rayet stars are known to be inhomogeneous, with 
clumps filling $\sim 1/4$ of the volume \cite{HK98}.
However, the observed
degree of inhomogeneity refers to a region close to the
stellar surface and the inhomogeneity may decrease in the outer parts
of the wind where the afterglow occurs. 

\subsection{Jets}

Jet breaks themselves provide a possible diagnostic for the
medium.
The usual assumption has been that an afterglow light curve steepens
due to geometric and spreading effects when its Lorentz factor is
about 1/(opening angle) to the form $t^{-p}$ \cite{SPH99}.
Kumar \& Panaitescu \cite{KP00} 
find that the transition to the asymptotic jet evolution
requires a factor $\ga 10$ in time for expansion in a uniform medium, but
expansion by $\sim 10^4$ in time for expansion in a wind medium.
However, they used a simplified treatment of jet evolution.
Numerical simulations indicate that sharp jet breaks do occur in a uniform
medium, but that most of the emission remains within the initial
opening angle of the jet \cite{Ge01}.
Granot \& Kumar \cite{GK03} have recently considered structured jets
in uniform and declining density media, and again found that jets
in an $s=2$ medium cannot give sharp jet breaks.

The recent burst GRB 030329, which was clearly associated with a supernova,
showed a sharp break in the light curve at $t\sim 0.5$ day, which
has been interpreted as a jet break \cite{Be03}.
However, it has become clear that there is considerable structure in
the light curve of GRB 030329 over the first 10 days \cite{Ue03,Ma03}
and the identification of an early jet break cannot be made with  certainty.

\subsection{Afterglow parameters}

There are uncertainties in the basic assumptions involved in standard
afterglow modeling, which include constant values of electron energy
efficiency, $\epsilon_e$, magnetic energy efficiency, $\epsilon_B$,
and particle spectral index, $p$, in the evolution of one source
(e.g., \cite{SPN98}).
If these parameters remain constant during the evolution of one source,
the expectation would be that they tend toward ``universal'' values
that apply to various sources.
Standard models developed for observed afterglows do not
show this.
As an example, I take the results of Panaitescu \& Kumar \cite{PK02}, 
who treat a set of the 10
best observed afterglows with the standard assumptions.
The values of spectral index, $p$, cover the range $1.36-2.78$.
The presence of values $p<2$ is noteworthy because most of the particle energy
is at high energy for this case, although the number of particles
is dominated by the low energy particles.
The values of $\epsilon_B$ and $\epsilon_e$ cover the ranges of $4\times
10^{-5}-0.07$ and
$0.01-0.4$, respectively.

The theoretical values of these parameters 
  are poorly known.
The production of the magnetic field requires some mechanism at the forward
shock front to build up the magnetic field.
The mechanism remains uncertain, although the Weibel instability has been
suggested \cite{ML99}, and recent simulations show some promise for
this mechanism \cite{Ne03,Fe03}.
Studies of Fermi-type 
particle acceleration in ultarelativistic shocks have yielded a
preferred value of $p=2.2-2.3$ in the test particle limit \cite{Ae01}.
The way electrons are injected into the acceleration process
remains uncertain, and there are indications that the acceleration
may differ from the Fermi process \cite{Ne03,Fe03}.

The fact that a range of parameters is needed to explain the various afterglows
suggests that the parameters depend on the physical conditions.
It might be expected that the shock velocity and preshock density are
important determinants of the physical conditions, so that the model parameters
should vary during the evolution of a burst.
Yost et al. \cite{Ye03} have recently considered models in which $\epsilon_B$ 
is allowed to vary as a power law of the shock Lorentz factor and in which
the value of the preshock density parameter $s$ is allowed to cover
a large range.
Their modeling of 4 sources
 shows that a wide variety of models are potentially possible,
including ones in which the density increases steeply with radius.
It appears that more extensive observations, including spectral
information over a wide time range, are needed to further constrain
the models.
One possibility is to follow the evolution of the characteristic
frequency, $\nu_m$, from optical to radio wavelengths.
This frequency is typically observed at radio wavelengths on a
timescale $\sim 10$ days.
Its passage at optical wavelengths requires very early observations;
as noted above, this may have been observed in GRB 021004.
The typical frequency is not sensitive to the density, but it is sensitive
to the efficiency factors, so that constraints on their evolution
may be obtainable.

\subsection{A shocked wind environment}

In view of the evidence for the association of GRBs with massive stars
and the evidence from afterglows for interaction with a constant
density medium,
consideration must be given as to how a massive star might produce a
uniform density surrounding.
The most plausible way to do this is the approximately constant density region
expected downstream from the wind termination shock \cite{W01}.
Wijers \cite{W01} suggested two ways for creating a smaller value of $t_t$ 
(eq. \ref{time}):
reducing the mass loss rate from the progenitor to $10^{-6}\ml$
because of the low metallicity of the progenitor star and increasing
the pressure by interacting with
dense molecular gas, especially if the progenitor star is moving.
However, the metallicity dependence of the mass loss from Wolf-Rayet stars
is uncertain: WN type stars in the lower metallicity Large Magellanic
Cloud \cite{HK00} and the Small Magellanic Cloud \cite{C00} have similar
mass loss rates to those in the Galaxy, although WC stars do seem to
show a metallicity effect \cite{Ce02}.
Also, 
photoionizing radiation during
the life of a massive star tends to clear a region around the star to
a moderately low density.
Another way of increasing $p$ is by having the burst occur in a high pressure
starburst region \cite{F03}, where the pressure can reach values of
$p/k\ga 10^8$ cm$^{-3}$ K \cite{CF01}.
In this case, there should be a relation between the properties of the
afterglow (relatively dense surroundings) and the position of the burst
relative to a region of very active star formation.

One expectation of the models with a termination shock  is that some bursts
should be observed to make a transition from an $s=2$ to an $s=0$ medium, with
a density jump between them.
There has been little evidence for such a transition.
The expectation for such a transition is that the light curve should evolve
to a flatter asymptotic decline after a jump in flux.
Wijers \cite{W01} mentioned GRB 970508 because it had
a bump in the optical light curve at an age of 1 day.
However, it did not show the expected flattening of the light curve.
A burst that  showed a steepening with a possible bump is GRB 030226.
Dai \& Wu \cite{DW03} suggested that the transition was due to the interaction
with a large density jump, 
which might occur at the contact
discontinuity between the shocked progenitor wind and the dense red
supergiant wind from a previous evolutionary phase.
In this picture, the steepening of the light curve is due to
the sideways expansion of the jet in the dense medium.
However, the data on GRB 030226 are not of sufficient quality to
clearly show the expected features at the time of transition with the
density jump.
In the case of interaction with the termination shock of the stellar
wind, the density jump is such that
the effects of the reverse shock are not expected to be important, as
opposed to the high density jump case.

\section{Optical/ultraviolet absorption lines}

A recent development relevant to the surroundings of
GRBs is the observation of strong optical/ultraviolet
 absorption lines in a few
cases.
The best case is GRB 021004, which has a redshift $z=2.32$ so
that strong ultraviolet  lines in the rest frame are redshifted
to optical wavelengths \cite{Me03,Sce03}.
Strong lines of Ly$\alpha$, Ly$\beta$, C IV, and Si IV are found
blueshifted relative to the host velocity by $-450$, $-990$, 
and $-3155\kms$ \cite{Me03}.
The lines have not been observed to vary, so they cannot be
directly tied to the immediate circumburst environment, but
both Mirabal et al. \cite{Me03} and Schaefer et al. \cite{Sce03}
argue that they are likely to be formed in the nearby environment.
One  argument for this is that the strong lines
are unusual for intervening systems observed in the spectra
of quasars.
The lines may be formed close to the host galaxy if the burst
occurred in a starburst region with a strong galactic superwind.
However, the maximum velocity shift is higher than has been
observed in galactic superwinds.
In addition, the high velocity would require a high initial 
temperature for the gas if the wind is thermally driven.
The gas would be completely ionized and it is unlikely that
it would be able to cool to allow the observed ions.

For a circumstellar origin, there are two possibilities:
the high velocities are related to the wind velocities of
the progenitor system, or the velocities are due to radiative
acceleration by the GRB light; combinations of these models
are also possible.
Schaefer et al. \cite{Sce03} 
argue that the high blueshifted velocity can be naturally
produced by the Wolf-Rayet star wind velocity and the lower
velocity components can be identified with denser shells
swept-up by the fast wind.
Mirabal et al. \cite{Me03} argue that the abundances deduced
from the lines disfavor the wind model.
Hydrogen is present in the observed lines, but it is also
observed in the spectra of some WN stars \cite{SSM96}; however, the
absence of N V lines in the observed spectra indicates that
N is not overabundant.
In the radiatively accelerated model, the accelerated clumps must be at
an initial distance of several 0.1's pc from the progenitor
star.
The acceleration, primarily by bound-free transitions, must
occur early in order to avoid the observation of time variability.

The problem with both of these scenarios is that the
strong radiation field from the GRB and its afterglow
is able to completely photoionize the gas out to a distance
$\ga 10^{18}$ cm \cite{Le02}.
At a sufficiently high density ($\ga 10^7$ cm$^{-3}$), the 
recombination time becomes shorter than the age of the burst.
This density might be present in clumps of the swept-up
red supergiant wind, especially if the progenitor is in
a high pressure region, but further exploration of this topic
is needed.

GRB 021004 is not alone in showing these line features.
High excitation, high velocity absorption
features have been found in GRB 020813 \cite{Ba03} and
GRB 030226 \cite{Ge03,Pr03,CF03}.
The absorption lines of CIV in GRB 020813 are at $0\kms$
and $-4320\kms$ relative to the host.
In this case, the blueshifted absorption is also present in a number
of lower ionization species (Si II, Al II, Fe II, Mg II, and Mg I);
there is no coverage of Ly$\alpha$.
In the case of GRB 030226, strong absorption line systems are present
at a velocity separation of $2300\kms$, with C IV and Si IV present, as well
as numerous lower ionization species and Ly$\alpha$.
The velocity separation seen in these sources is consistent with
expectations for the velocity of a Wolf-Rayet star wind.
However, the presence of H does not support this origin for
the lines.

\section{Pulsar wind bubble environment}

An interesting possibility for a GRB environment is that created by
a pulsar wind nebula.
This possibility was   proposed in the context of the supranova model
in which the supernova precedes the GRB \cite{VS98,KG02}.
In this scenario, the supernova core contracts to a massive, rapidly rotating 
neutron star which spins down and collapses to a black hole after a period
of weeks, months, or years.
If the neutron star has a magnetic field similar to that of
radio pulsars, it can create a shocked bubble
of relativistic electron/positron fluid and magnetic field before it collapses.
The bubble accelerates the supernova ejecta, so that the ejecta can play
a role in producing the X-ray lines that have possibly been observed
in some bursts.

One issue is how effectively the pulsar nebula can accelerate the
supernova gas, because the situation is subject to Rayleigh-Taylor
instabilities \cite{KG02,IGP03}.
The X-ray line features are typically blueshifted by $\sim 0.1c$, which is higher
than the velocities of the heavy element ejecta that would be expected
from the supernova itself.
A supernova energy of $10^{51}$ ergs in an ejecta mass of $10\Msun$
leads to a typical velocity of $0.01c$.
K\"onigl \& Granot \cite{KG02} suggest that the rotational energy
of a rapidly rotating neutron star, $10^{53}$ ergs, is transferred
to the ejecta, giving the observed velocity.
Although a pulsar nebula can certainly shock and compress the ejecta
gas, the ability to further accelerate the ejecta is less certain.
However, if the supernova 
explosion energy is high ($>10^{52}$ ergs as inferred
for some supernovae) and the ejecta mass is low, a typical velocity
of $0.1c$ can be attained.

Another issue is the fact that the observation of X-ray line features
at an age of $\sim 1$ day requires dense gas at $r\sim 10^{16}$ cm,
but observations of afterglow emission at an age of a week or more
imply a radial scale $\ga 3\times 10^{17}$ cm.
K\"onigl \& Granot \cite{KG02} suggest that pulsar nebula and supernova
may be highly elongated along the axis along which the GRB flow
propagates.

Although the observational evidence for X-ray lines remains controversial,
there are other reasons for considering a pulsar wind nebula environment,
as articulated by K\"onigl \& Granot \cite{KG02}. 
One is that the pulsar nebula is composed of just the ingredients
that are necessary for the synchrotron emission from a GRB afterglow.
There is no problem with the efficiencies for production of the
synchrotron emission.
In addition, the bubble density can be constant with radius, or drop with
radius.   
The first case can occur in the same way that a constant density occurs
downstream from an ordinary stellar wind.
The decreasing density occurs in regions where the  magnetic
field pressure becomes important, although the structure of such
regions in pulsar nebulae remains uncertain.

Although the pulsar bubble model has some appeal,  it does not apply to cases
where the GRB occurs close in time to the supernova, as apparently
was the case with GRB 030329 and SN 2003dh (e.g., \cite{Ma03}).

\section{Discussion and conclusions}

Despite several years of effort, the study of GRB
environments from their interaction has not clearly
pointed to the progenitor objects.
Reasons for this include the uncertainties in the
basic model parameters, the fact that the GRB ejecta
appear to be collimated and the similarity in the
surrounding densities expected in different scenarios.
The clearest progenitor information we have comes from
the association of SN 2003dh with GRB 030329.
The similarity of the supernova to SN 1998bw suggests that
the progenitor object is a similar massive star.
Analysis of the radio emission from SN 1998bw showed
compatibility with expectations of a wind-like surrounding
medium \cite{LC99}.
The radio observations of GRB 030329 are slightly better fit
by a uniform medium than a wind-like medium, although the
difference between the fits is not large \cite{Be03}.
The host galaxy of GRB 030329 appears to be a starburst
dwarf galaxy \cite{Ma03}
and the burst is positioned near the edge of
the star forming region \cite{Fue03}, so it is not clear
whether a high pressure surroundings is expected.
Detailed modeling will be needed to determine whether the 
afterglow features can be explained by a complex explosion,
or whether the surroundings need special properties.

An outstanding question is whether there is any need for
a progenitor of long-duration bursts other than massive stars.
Since massive stars are expected to modify their surroundings
through winds, this evidence would be an incompatibility
with the wind effects expected around a massive star.
Frail et al. \cite{Fr03} and Yost et al. \cite{Ye03} have noted
that a number of afterglows that can best be modeled as expanding
into a uniform density medium with $n\sim 10-30$ cm$^{-3}$,
and that this density is typical of Galactic interstellar clouds
or the interclump medium of molecular clouds.
However, such a medium would be modified by the winds from a massive
star progenitor and a significant fraction of GRBs probably occur
in starburst regions where clouds are denser than in Galactic case.
One possibility is that the uniform medium is created by a shocked stellar
wind in a high pressure medium.

A possible problem for massive star models is the low density
inferred around some GRBs.
Even if a shocked wind is present, it has higher density at a given
radius than the corresponding free wind, so strong limits on
the wind density can be set.
Two afterglows with low densities are GRB 990123 \cite{PK02}
and GRB 021211 \cite{KP03}, which require $A_*\la 10^{-3}-10^{-2}$.
Such low densities have not observed around Wolf-Rayet stars.
GRB 021211 shows evidence for a supernova \cite{DVe03}, but the
evidence is not conclusive.
The low densities are not a problem for a burst that interacts
directly with the hot interstellar medium in a galaxy.
Another problem with the massive star models is the lack of evidence
for bursts crossing the interface between a free wind and a
shocked wind.

If direct interaction with the interstellar medium is required,
a plausible progenitor object is a binary of compact objects.
This requires that two different progenitor objects can give
GRBs that appear similar, presumably from the formation of
black holes.
The application of the pulsar wind nebula model to some bursts
requires both that different progenitor objects give rise to
similar bursts, but also that the expansion of the burst into a
different kind of medium (pair plasma and magnetic field) can
give rise to similar afterglows.
The finding of a supernova (SN 2003dh) occurring at 
approximately the same time
as a GRB (GRB 030329) mitigates against the pulsar wind nebula
picture for this case.

\medskip

I am grateful to Jon Marcaide and the other conference organizers
for their hospitality in Valencia, and to Zhi-Yun Li and Claes
Fransson for continuing collaboration on these topics.
This work was supported in part by NASA grant NAG5-13272
and NSF grant AST-0307366.
%
%

%
%

%


\printindex
\end{document}